\documentclass[1p,preprint]{elsarticle}

\usepackage[utf8x]{inputenc}
\usepackage{graphicx}
\usepackage{amsmath}
\usepackage{amsfonts,color}

\def\U{U^{(1)}} 
\def\UU{U^{(2)}} 
\def\V{V^{(1)}} 
\def\VV{V^{(2)}}
\def\A{A^{(1)}} 
\def\F{F_{nl}} 
\newcommand{\dd}[1]{\frac{{\mathrm d}#1}{{\mathrm d}t}}
\newcommand{\DD}[1]{\frac{{\mathrm d}#1}{{\mathrm d}T}}
\begin{document}
\begin{frontmatter}

\title{Mechanisms for transient localization \\ 
in a diatomic nonlinear chain}
\author[isc,infn]{Stefano Lepri}
\address[isc]{Consiglio Nazionale delle Ricerche, Istituto dei Sistemi Complessi, Via Madonna del Piano 10 I-50019 Sesto Fiorentino, Italy} 
\address[infn]{Istituto Nazionale di Fisica Nucleare, Sezione di Firenze, 
via G. Sansone 1 I-50019, Sesto Fiorentino, Italy}
\author[cbm]{Francesco Piazza}
\address[cbm]{Centre de Biophysique Mol\'eculaire, (CBM), CNRS-UPR 4301, Rue C. Sadron, 45071, Orl\'eans, France
and Universit\'e d'Orl\'eans, Ch\^ateau de la Source, 45071 Orl\'eans Cedex, France}
\begin{abstract}
We investigate transient nonlinear localization, namely
the self-excitation of energy bursts in an atomic lattice
at finite temperature.
As a basic model we consider the 
diatomic Lennard-Jones chain.
Numerical simulations suggest that the effect  
originates from two different mechanisms. One is the
thermal excitation of genuine discrete breathers with 
frequency in the phonon gap. The second is an
effect of nonlinear coupling of fast, lighter particles with slow
vibrations of the heavier ones. The quadratic term of 
the force generate an effective  potential 
that can lead to transient grow of local energy on time 
scales the can be relatively long for small mass ratios.
This heuristics is supported
by a multiple-scale approximation based on the natural
time-scale separation. For illustration, we 
consider a simplified single-particle model that 
allows for some insight of the localization dynamics.
\end{abstract}
\end{frontmatter}

\section{Introduction}

Energy transfer among nonlinear systems occurs in many physical
context, ranging from condensed matter to optics. For instance,   
understanding the principles of vibrational energy transport at the nanoscale
is a necessary step for thermal management in phononic systems and
requires a deeper understanding of the properties of strongly anharmonic
and/or disordered crystals and artificial materials.
Nonlinear effects are essential in many respects: in the 
first place, they determine thermal
transport properties. This is particularly dramatic in low-dimensions 
where nonlinear interaction of energy fluctuations lead to anomalous 
conductivity \cite{Lepri2016}. Also, nonlinear excitations have proven
to be responsible of slow-relaxation phenomena, whose dynamics recalls
that of glassy systems despite the fact that disorder is not present 
(see e.g. \cite{tsironis1996slow,piazza2001slow,rumpf2004simple,iubini2013discrete,iubini2019dynamical} and references therein). There is also evidence that self-excitation of DB plays a role 
in nonequilibrium steady states \cite{iubini2017}.

The concept of localization due to nonlinearity is well established.
For discrete nonlinear systems, 
the notion of discrete  breathers (DB), termed also intrinsic localized modes (ILM) in solid state physics or
discrete solitons (DS) in nonlinear optics, is well known. DB are exact time-periodic and spatially localized vibrational modes found generically in nonlinear lattice models, typically independent of the size and 
dimensionality of the lattice, and of the specific choice of anharmonic potentials. Their existence emerges as a joint effect of anharmonicity (i.e. an energy-dependent frequency of vibration) and discreteness (i.e. the existence of band gaps in the plane-wave spectrum) \cite{flach2008discrete}.

A relevant problem is the self-excitation of such localized modes 
in thermodynamic conditions. This has been studied   in the literature
for several models \cite{kosevich2000modulational,
eleftheriou2003breathers,ivanchenko2004discrete,Farago20081013}. 
For anharmonic lattices at thermal equilibrium (i.e. at finite energy 
density) one may expect that DBs may be created as a kind of randomly activated
process. Relatively large thermal fluctuations may lead to shift of
the local oscillation frequency, that may enter the plane-wave spectral gap. Once formed, 
such localized modes may stay off-resonance from the linear spectrum 
allowing the energy to remain confined at a few sites for relatively 
long times. This manifest as a form of \textit{transient localization}. However, it is not trivial to identify 
unambigously the signature of DB directly from equilibrium
simulations, i.e. by direct inspection of particle trajectory.
Some specific diagnostics has been indeed proposed
both in the frequency and time domains \citep{forinash1998frequency,eleftheriou2005discrete,Farago20081013,PhysRevB.84.144304}.

In this work we will address this problem both numerically than
analytically for a simple diatomic chain. We argue that transient localization observed empirically 
originates from two different mechanisms. One is the
thermal excitation of genuine DB with 
frequency in the phonon gap. The second is an
effect of nonlinear coupling of fast, lighter particles with slow
vibration due to the heavy ones. We illustrate our findings by
considering 
a simple model consisting of particles constrained on the line and interacting with a nearest-neighbor Lennard-Jones potential with diatomic arrangement of 
masses.  The uniform (equal-masses) case  has been considered in several papers
starting from ref. \cite{Mareschal88}. A relevant
result is that the dynamical correlations function  
display breakdown of conventional hydrodynamics. This anomaly is traced back by the presence of correlations due to the reduced dimensionality \cite{Lepri2005}. 
The existence of DB is proved for alternating mass chains with anharmonic coupling for large enough mass ratio was first proved in \cite{livi1997breathers} and later studied in several works, 
see e.g \citep{maniadis2003existence,gorbach2003discrete,zolotaryuk2001discrete,james2007bifurcations,boechler2010discrete}.  
Besides this, diatomic chains have been studied in the non-equilibrium setup for the alternating mass harmonic \cite{kannan2012} and 
anharmonic \cite{hatano1999heat} cases. The chain with alternating bonds
has also been considered \cite{Xiong2013}.

The outline of the paper is as follows. In Section \ref{sec:model}
we present the model and recall its harmonic approximation in 
Section \ref{sec:harmo}. Numerical results 
based on power spectra and wavelet transforms
are presented in Section \ref{sec:loc}.
In Section \ref{sec:mult} we present an analysis of the 
case where the mass ratio is small and nonlinearity weak.
A novel type of multiple scale expansion is presented 
yielding approximate equations for the motion of 
light particles. A simplified single-particle dynamics
is illustrated to explain heuristically the transient 
energy localization mechanism. Finally, a brief 
summary is given in the last Section.

\section{The diatomic Lennard-Jones chain}
\label{sec:model}

We consider an array of $N$ point-like atoms ordered along a line. 
The position of the $n$-th atom is denoted with $x_n$ and let $m_n$
denote its mass. Assuming that interactions are
restricted to nearest-neighbor pairs, the equations of motion write
\begin{equation} 
m_n{\ddot x}_n = V'(x_{n+1} - x_n) - V'(x_{n} - x_{n-1})\, ,
\label{eqmot} 
\end{equation}  
where $V'(z)$ is a shorthand notation for the first derivative of the
the interparticle potential $V$ with respect to $z$. The particles 
are confined in a simulation ``box"
of length $L$ with periodic boundary conditions 
\begin{equation}
x_{n+N} \;=\; x_{n} \,+\, L  \quad .
\label{pbc}
\end{equation}
Accordingly, the particle density $d=N/L$ is a state variable to be considered
together with the specific energy (energy per particle) that will be denoted by
$e$. We focus on the Lennard--Jones potential
that in our units reads \cite{Lepri2005}
\begin{equation}
V(z) \;=\;  {1\over 12}
\bigg({1 \over z^{12}}\, -\, {2\over  z^6}\,+\, 1 \bigg)\quad .
\label{lj}
\end{equation}
For computational purposes, the coupling parameters have been fixed in such a
way as to yield the simplest form for the force. With this choice, $V$ has a
minimum in $z=1$
and the resulting dissociation energy is $V_0=1/12$. Notice that for
convenience we set the zero of the potential energy in $z=1$.  The presence of
the repulsive term in one dimension ensures that the  ordering is preserved
(the particles do not cross each other).  

We consider a diatomic chain with $m_n$ assuming two alternating values $m$ and $M$. Before proceeding, we recall that 
Huang and Hu \cite{PhysRevB.57.5746} argued that the potential (\ref{lj})
satisfies the condition for the existence of optical gap breathers 
(that they call optical lower cut-off gap soliton modes) if the 
cubic term in the Taylor expansion is large enough 
but does not admit acoustic gap breathers nor optical 
above the optical band. However, acoustic and optical upper cutoff vibrating kinks are possible excitations for these diatomic lattice systems.
According to their paper this is true independently of the mass ratio.
{This property of the diatomic Lennard-Jones potential
should be contrasted with other cases like with hard nonlinearity where 
DB close to the acoustic band can form, see e.g. \citep{Xiong2013}.}
In \cite{hu2000dynamics} the possibility of more exotic  
kink-solitons solutions is also demonstrated. 

\section{Harmonic approximation}
\label{sec:harmo}
From now on, we choose to work at a density value $d=1$ 
such that the equilibrium positions of the particles coincide
with the mimimum of the potential $V$.
It is convenient to
separate the equations of motion for even-numbered (light) and odd-numbered (heavy) particles
introducing the  displacements $u_n,v_n$ from equilibrium
positions {(that in our dimensionless units can
be set to $nl$ with $l=1$)} as
$x_{2n+1}=2n+1+u_n$ (slow) and $x_{2n}=2n+v_n$ (fast) \cite{porubov2013nonlinear}. They
satisfy the equations  ($F\equiv -V'$, $F(0)=0$)
\begin{eqnarray}
&&m{\ddot v}_{n}= - F(u_{n} - v_{n}) + F( v_{n} - u_{n-1})\quad ; \qquad \\
&&M{\ddot u}_{n}= - F(v_{n+1} - u_{n}) + F( u_{n} - v_{n}) \nonumber
\end{eqnarray}
For later reference it is useful to recall the harmonic appproximation
of the chain where $F(z)\approx -kz$
with $k=6$ in our units for $V$ as given by eq. (\ref{lj}).  The phonon bands $\omega(q)$ 
for the infinite chain are
\begin{equation}
 \omega^2(q)=k\left(\frac{1}{M}+ \frac{1}{m} \pm\
 \sqrt{\left(\frac{1}{m}+\frac{1}{M}\right)^2-\frac{4}{mM}\sin^2(q)}\right)
 \label{phonons}
\end{equation}
and $|q|<\pi/2$ is the Brillouin zone.
There is a band gap in the linear spectrum defined by the band edges 
of the acoustic and optical branches:
\begin{equation}
 0<\omega < \sqrt{\frac{2k}{M}},\qquad \sqrt{\frac{2k}{m}} <\omega <\sqrt{2k(\frac{1}{m}+\frac{1}{M})}.
 \label{banded}
\end{equation}
The sound velocity defined in the small $q$ limit of the acoustic branch
is another relevant scale of the model and is given by
\begin{equation}
v_s = \sqrt{\frac{2k}{m+M}}
\end{equation}

\section{Transient localization}
\label{sec:loc}
In this section we first illustrate the results of numerical data 
demonstrating transient localization  in the harmonic
gap defined by equations (\ref{banded}).
We have performed equilibrium microcanonical simulations by integrating
Eqs.~(\ref{eqmot}) (with periodic boundary conditions (\ref{pbc}) )  by means
of a fourth--order symplectic algorithm \cite{mclachlan1992accuracy}. Initial 
positions were chosen to be in the
ground state as given in the  previous Section. The initial velocities were
drawn at random  from a Gaussian distribution and rescaled by suitable factors
to assign the kinetic energy to the desired value and to set the total initial
momentum equal to zero. A suitable transient is elapsed before acquisition of 
statistical averages. Conservation of energy and momentum was monitored during
each run. This check is particularly crucial at high energies/densities where
the strongly repulsive  part of the force comes into play and may lead to
significant inaccuracies. The chosen time--step (0.01) ensures
energy conservation up to a few parts per  million in the worst case.   

\begin{figure}[htp]
 \begin{center}
 \includegraphics[width=0.45\textwidth,clip]{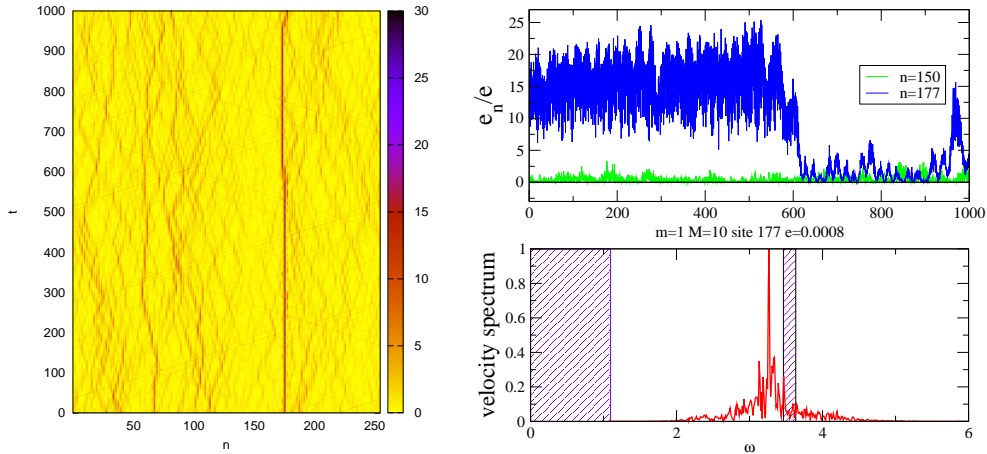}
 \includegraphics[width=0.5\textwidth,clip]{fftsite177.eps}
 \caption{Transient localization of energy in the diatomic Lennard-Jones chain
$N=256$, $d=1$, $m=1,M=10$. 
Left: space-time plots of the local energies $e_n$.
Right upper panel: comparison of time evolutions
of energies on the breather site with the one without. 
Right-lower panel: Fourier spectrum of the 
momentum of the particle $n=177$, showing the 
emergence of nonlinear oscillations in the 
harmonic gap (the shaded area represent the two linear bands
given by eq.(\ref{banded}).}
 \label{fig:f1}
\end{center}
\end{figure}

In Fig. \ref{fig:f1} we show the space-time plots of the 
local energy 
\[
e_n= \frac12 m_n\dot{x_n}^2 + \frac12[V(x_{n+1} - x_n) + V(x_{n} - x_{n-1})]
\]
of one representative run. There is a spontaneous localization 
of energy around well-defined sites where an energy above
the average it is seen for a given time. In the rightmost 
panels the time evolution of local energy and the 
associated power spectra of the momentum are illustrated.
The characteristic 
oscillation frequency lies in the gap of the linear spectrum
just below the optical band, 
thus supporting the idea that it correspond to excitation 
of DB with a given lifetime. This concept has been discussed
often in the literature for various nonlinear models.
From the figure it is nonetheless evident that also 
other types of localized spots appear throughout the lattice.


A convenient way to analyze non-stationary signals is to use a wavelet analysis in the time-frequency domain.  This technique proved to be useful
to pinpoint transient vibrational excitations in many-body
system starting from simple 
chain models \cite{forinash1998frequency}
to simulated NaI crystals \cite{riviere2019wavelet}.
This method allows to detect transient frequency components appearing at specific times and lasting for finite lapses of time.  
In this work, we have computed the Gabor transform of 
the momentum of  $n$th oscillator, namely
\begin{equation}
\label{e:gabor}
G_n(\omega,t) = \int_{-\infty}^{+\infty} e^{-(t-\tau)^2/a} e^{-i\omega \tau } p_n(\tau ) \, d\tau 
\end{equation}

\begin{figure}[htp]
 \begin{center}
  \includegraphics[width=0.9\textwidth,clip]{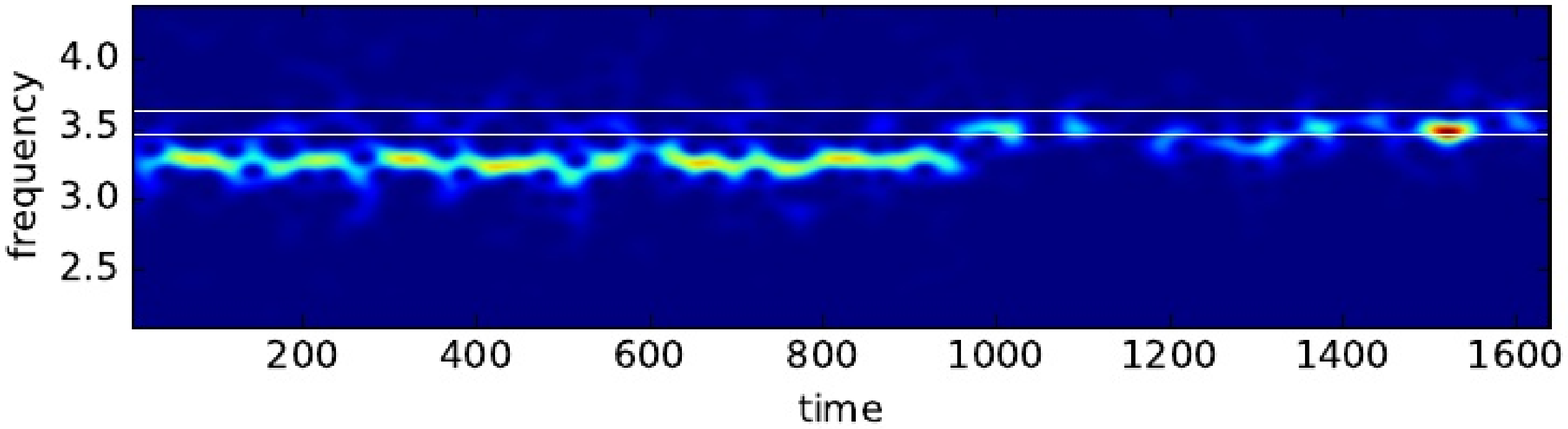}
 \includegraphics[width=0.9\textwidth,clip]{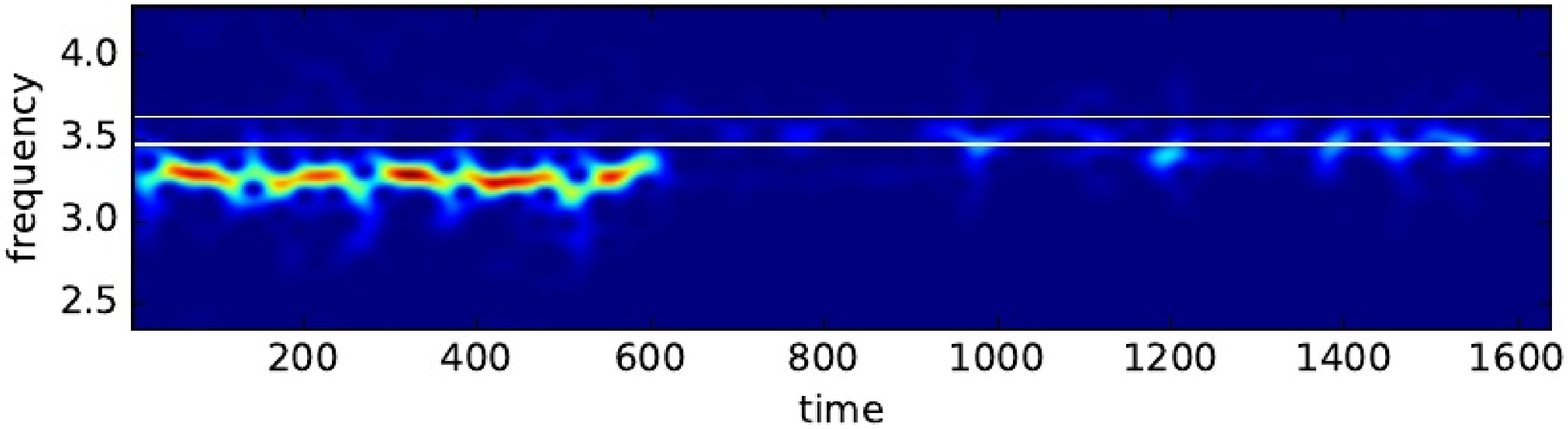}
 \includegraphics[width=0.9\textwidth,clip]{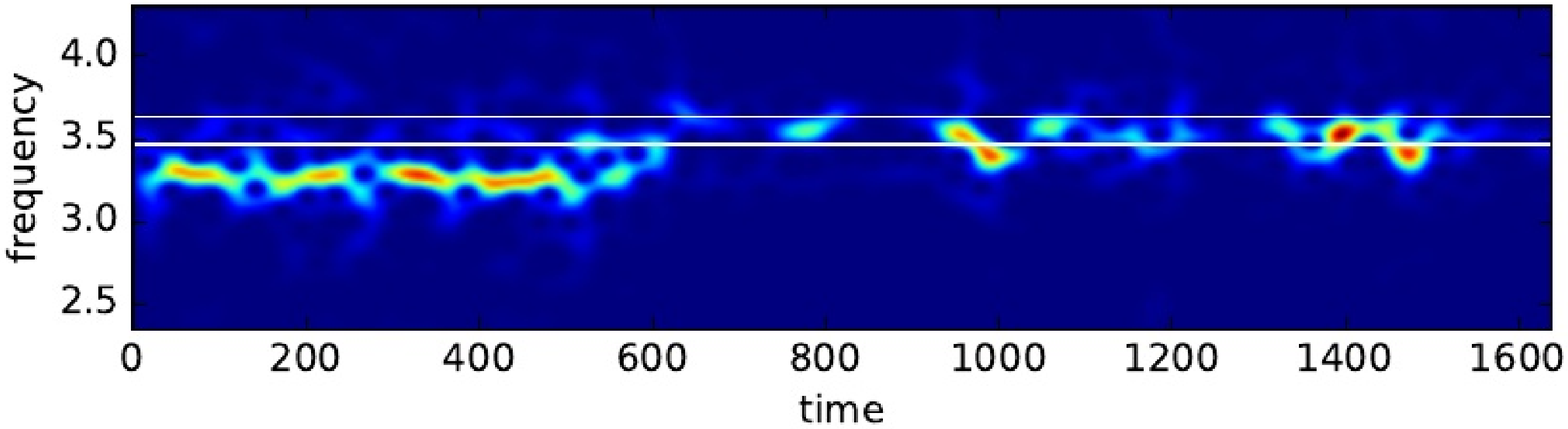}

 \caption{The square modulus of 
 the Gabor transform  of momentum for light and heavy particles is different, $N=256$, $d=1$, $m=1,M=10$,$a=500$ for three adjacent sites $n=176,177,178$ (top to bottom). The horizontal lines correspond to the edges 
 of the optical band. Even sites have 
lighter mass $m$ and display shorter transient localization
slightly below the lower optical band-edge. The signature of the oscillations
at a breather frequency 3.3 is seen in all of the neighboring 
particles. }
 \label{fig:wave}
\end{center}
\end{figure}

In Fig.\ref{fig:wave} we show $|G_n(\omega,t)|^2$ for three adjacent 
sites for the same run as in Fig.\ref{fig:f1}. First of all 
there is a clear confirmation of the DB oscillations below the optical
band. Notice that the signal is present simultaneously 
on all the three sites in agreement with the idea that the 
DB is a collective oscillation that involve all the 
particles, independent of their mass.
Besides that there are signatures of transient localization
close or below the lower acoustic band edge. This type of
events seems to occur independently on neighboring sites.

Besides considering individual trajectories one may look at 
statistical indicators like correlation functions.
We computed the dynamical structure factor, namely the square modulus of 
temporal Fourier transform of the particle density 
\begin{equation}
\rho(q,t)  \;=\; {1\over N}\,\sum_n \, \exp(-iqx_n) \quad ,
\label{dens}
\end{equation}
which is defined as
\begin{equation}
S(q,\omega) \;=\; 
\big\langle \big| \rho(q,\omega )\big|^2 \big\rangle  \quad .
\label{strutf}
\end{equation}
The square brackets denote an average over a set of independent
molecular--dynamics runs. By virtue of the periodic boundaries, the allowed
values of the  wavenumber $q$ are integer multiples of $2\pi/L$.
Data windowing (Hanning window) has been used to compute the FFT in time.
We also computed the spectrum of the momenta $p_n=m_n \dot x_n$ of the individual particles
\begin{equation}
 s_n(\omega) = \langle \big| p_n(\omega )\big|^2 \rangle 
 \label{inco}
\end{equation}
averaging over independent trajectories.   This 
quantity is related to the incoherent part of the spectrum 
according to the terminology used in neutron scattering experiments.

\begin{figure}[htp]
\begin{center}
 \includegraphics[width=0.8\textwidth,clip]{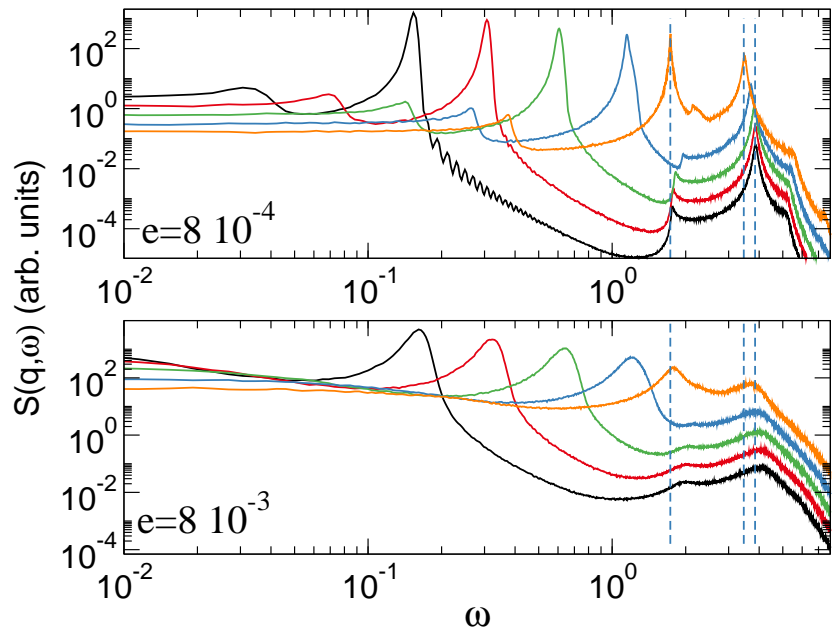}
 \caption{Structure functions for $q= 0.0981, 0.1963, 0.3926, 0.7853 $ and $\pi/2$ from left to rigth;
 $N=1024$, $d=1$, $M=4$, $m=1$
 the vertical dotted lines are the phonon band edges.
 Two values of the energy density $e$ are shown. For larger 
 $e$ the spectral components in the gap is larger.}
 \label{fig:sfun}
 \end{center}
\end{figure}

In Fig. \ref{fig:sfun} we plot the structure factors for two
different values of the energy density $e$ and different wavenumbers
$q$. The main peaks are in good agreement with the phonon frequencies 
calculated from eq. (\ref{phonons}). What appears is that, upon increasing
the energy density there is an increasingly larger spectral component
within the band-gap, that signals the enhancement of 
gap oscillations due to nonlinearity. Moreover, the largest frequency 
component in the gap occurs for wavenumbers closer to 
the zone boundary, $q\approx \pi/2$ meaning that the nonlinear 
effects and transient localization is, as expected, associated
to short-wavelengths dynamics.

\begin{figure}[htp]
 \begin{center}
 \includegraphics[width=0.85\textwidth,clip]{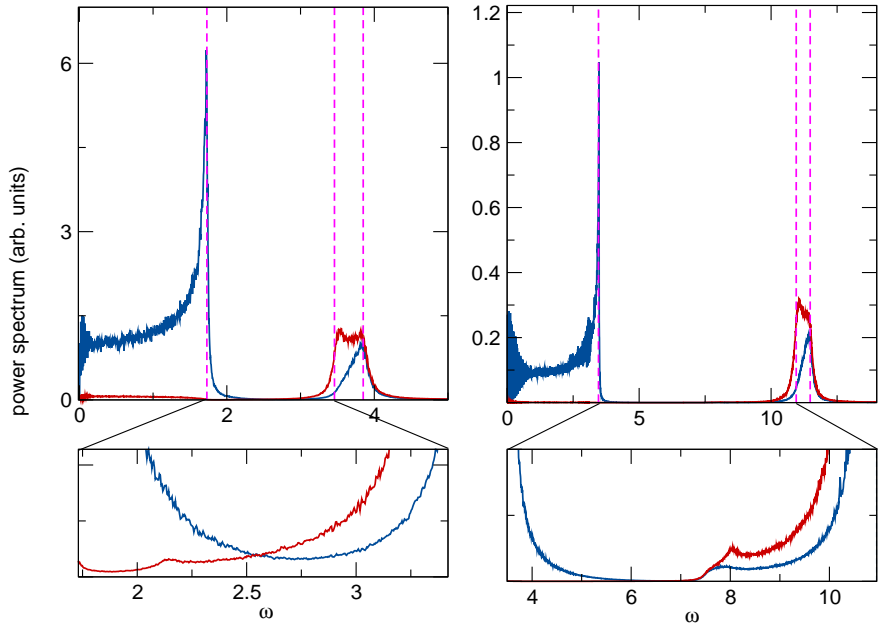}
 \caption{The spectra of momentum for light (red) and heavy (blue) particles is different, $N=1024$, $d=1$ energy density $e=8.0\,\cdot 10^{-4}$. 
Left panels: mass ratio $m=1$,$M=4$, right  $m=0.1$,$M=1$.
The vertical dotted lines are the phonon band edges. The bottom
panels are enlargements of the gap regions.
{Notice how in the bottom-right panel the small peak 
at $\omega \approx 8$ is only
present in the spectrum of light masses: this can be interpreted 
as due to localized oscillations with almost no component
on the heavy particles}.}
  \end{center}
   \label{fig:si}
\end{figure}

In Fig.\ref{fig:si} we compare the spectra $s_n$  of light and heavy 
particles, as given
by definition (\ref{inco}), for two different mass ratios. To 
improve statistics an averaging over a subset of about 10 
particles with the same mass is performed. 
As expected from the 
harmonic approximation, it is seen that light particles 
oscillate faster with frequencies around the optical band
and a weak spectral component in the acoustic band.
Such a component reduces upon increasing the mass ratio.
Moreover, there is an increasing sizeable component in the gap that can be 
attributed to the localized excitations described above. 

The simulations suggest that there is a separation of time scales
in which the fast dynamics is driven by the slow motion
of the heavy masses. One may thus argue about
the consequences of such driving on transient energy localization.
This will be discussed in the next section
by means of a multi-scale approach.

\section{Approximate dynamics}
\label{sec:mult}

The first step is to distinguish fast and slow time-scales. 
Let us first introduce the smallness parameter $\varepsilon=\sqrt{m/M}\ll 1$.
To lowest order, the harmonic bands are thus given by
\begin{eqnarray} 
&&  \omega \approx \sqrt{2k} \varepsilon |\sin q| \\
&&  \omega \approx \sqrt{2k} \left(1+\frac{\varepsilon^2}{2} \cos^2 q\right)
\label{appphon}
\end{eqnarray}
In this limit the acoustic band is of order $\omega_0\varepsilon $
and the optical is in between $\omega_0$ and $\omega_0(1+\varepsilon^2/2)$
where $\omega_0\equiv \sqrt{2k}$.
The equation
of motion are
\begin{eqnarray}
&&m{\ddot v}_{n}= - F(u_{n} - v_{n}) + F( v_{n} - u_{n-1})\quad ; \qquad \\
&&m{\ddot u}_{n}= \varepsilon^2[- F(v_{n+1} - u_{n}) + F( u_{n} - v_{n})]
\label{eqmo}
\end{eqnarray}
that explicitly show how the separation of time scales occurring for $\varepsilon \to 0$. Already at this level, it is clear that the first eq. (\ref{eqmo}) 
is an equation for the oscillator $v_n$ subject to an effective force
changing slowly with the variables $u_n,u_{n-1}$. So it is 
in principle possible that such a force may destabilizes the oscillator,
at least for a 
certain time interval. This can be seen for instance for 
$v_n\ll u_n,u_{n-1}$. On the rhs of the eq. (\ref{eqmo}) 
for $m{\ddot v}_{n} $
there appears a term $\approx (F'(u_n)+F'(-u_{n-1}))v_n$ 
that can lead to transient growth of the energy if $F'(u_n)+F'(-u_{n-1})>0$
for a sufficiently large time lapse. A rough estimate of such time scale
is given by the inverse of the acoustic band edge, that from (\ref{appphon}) is
of order $\varepsilon^{-1}$. Altogether, such heuristic argument suggests that
a transient localization of energy can be seen on a relatively 
long time interval for $\varepsilon$ small enough.

The argument can be made more precise by a multiple-scale analysis
for weak nonlinearity.
This amounts to an expansion of the coordinates of the form
\[
u_n =  U_n+\varepsilon \U _n  +\varepsilon^2 \UU _n + \ldots 
;\quad v_n=V_n + \varepsilon \V _n  +\varepsilon^2 \VV _n + \ldots
\]  
where the $U$ and $V$ are a priori all functions of the time-scales 
$(t,T=\varepsilon t)$.
The details of the calculation are given in the Appendix. 
Performing the calculation up to order $\varepsilon^2$ 
and under some simplifying assumption, 
a closed set of equation can be obtained see eq. (\ref{a1})
in the Appendix.

A detailed analysis of the resulting equations will be reported
elsewhere. Here we limit ourselves to some qualitative considerations.
Indeed, a useful insight
can be achieved by considering the following equations 
\begin{eqnarray}
\label{oscmod}
 (i-\varepsilon)\Omega_0\DD{ \mathcal{A}_n} &=& 
 -\frac32 \beta |\mathcal{A}_n|^2 \mathcal{A}_n -\alpha\left(U_n-U_{n-1}\right)\mathcal{A}_n \\
 {\frac{{\mathrm d^2}U_n}{{\mathrm d}T^2}}
  &=& \frac12\omega_0^2[U_{n+1} - 2U_{n} + U_{n-1}]\nonumber
\end{eqnarray}
where the complex variable 
$\mathcal{A}_n$ represents the slow modulation of the 
amplitude of the light particle 
\begin{equation}
v_n(t,T)=  \frac {U_n+U_{n-1}}{2} + \mathcal{A}_n(T)  e^{i\Omega_0 t} + 
\mathcal{A}^*_n(T)  e^{-i\Omega_0 t}
\label{vslow}
\end{equation}
(this last equation approximates the sum of eqs. (\ref{ord0}) and 
(\ref{ord1}) below). This model can be regarded as 
a crude approximation
where the coupling terms in eq. (\ref{a1}) below and 
the fast component $\UU_n$are ignored.

Despite the underlying simplifications, 
the first equation in (\ref{oscmod}) is insightful:  it shows that the fluctuations
of the slow field appear multiplicatively and 
can be regarded as a dissipation or gain term 
depending on the sign of the local stretch $U_{n}-U_{n-1}$.
At finite temperatures, such a quantity will be an incoherent
superposition of all the harmonic modes of the harmonic
chain in (\ref{oscmod}). Thus, it is basically 
a kind of slowly-varying noise with a finite 
bandwidth driving the light particles. 

\begin{figure}[htp]
 \begin{center}
 \includegraphics[width=0.9\textwidth,clip]{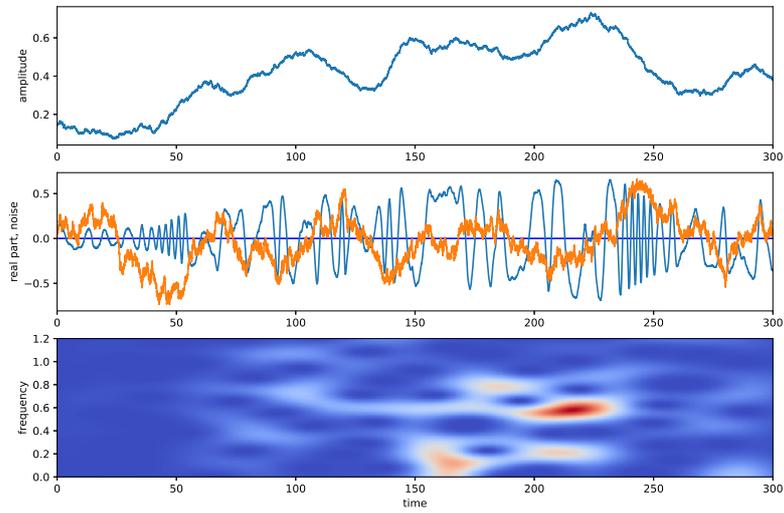}
 \caption{Langevin simulation of equation for $\mathcal A_n$ as in eq. (\ref{oscmod})
 where $U_n-U_{n-1}$ is replaced by an Ornstein-Uhlenbeck process $z$
 with correlation time $\varepsilon^{-1}$ and variance 0.1.
 $\varepsilon=0.025$, $\alpha=4\Omega_0$,$\beta=2\Omega_0/3$.
 Top: oscillator amplitude $|\mathcal A_n|^2$. Middle panel: real part
 of $\mathcal A_n$ and the process $z$; Bottom: the square modulus 
 of the Gabor transform of $\mathcal A_n$ displays transient 
 oscillations in correspondence with the instantaneous growth of the 
 oscillator amplitude.}
  \end{center}
   \label{fig:lang}
\end{figure}

Following this idea,
in Fig. \ref{fig:lang} we report 
a simulation of  Langevin simulation where the equation for $\mathcal A_n$ 
is as in eq. (\ref{oscmod}) while $U_n-U_{n-1}$ is replaced by an Ornstein-Uhlenbeck process $z$, $\dot z= -\varepsilon z + \xi$. Here, $\xi$ is a random Gaussian process with standard deviation $\sigma$. The random variable $z$
has a bandwidth of size $O(\varepsilon)$ that mimics  
the dynamics of $U_n$ on the slow scale. As seen in the bottom 
panel of Fig. \ref{fig:lang}, the wavelet analysis reproduces qualitatively the transient localization in frequency generated by the nonlinear frequency shift.
We thus conclude that our interpretation is supported by the effective model.

Finally, we may discuss the possible role of additional 
damping on the transient localization. If an additional small dissipation 
is added which is of the same order of $\varepsilon$, the dynamics 
should be qualitatively the same. Within the 
limitation of the simplified model (\ref{oscmod}), we may expect
the local instability mechanism should be robust, albeit slightly
inhibited by a larger dissipation level.

%
%

\section{Conclusions}

We have argued that transient nonlinear localization in the 
gap of the diatomic Lennard-Jones chain at finite temperature 
originates from two different mechanism. One is the
thermal excitation of genuine DB. The second is an
effect of coupling of the light particles with slow
vibrations of the heavy ones. The quadratic term of 
the force generate an effective  potential 
that can lead to transient grow of local energy on time 
scales the can be relatively long for small mass ratios.
As a consequence, the spots of localized 
energy created in this way are different in nature from 
thermally generated breathers. Thus, some caution is needed 
in the interpretation of transient localization event. 
This is a novel issue that, in our view, should be considered in the data interpretation.

The heuristics and numerical observations are supported
by a multiple-scale calculation based on the natural
time-scale separation. As a further simplification, we 
considered an effective single-particle model that 
allows for some insight of the chain dynamics.
We remark that multiple scale approach used here 
is different from the standard one employed to 
study DB solutions based on envelope instability of 
zone-boundary modes \citep{flach2008discrete}. 
Here we rather consider the question of how the fast 
dynamics is affected by slow motion and should thus 
be regarded as a complementary approach.

\section*{Acknowledgements}
This research did not receive any specific grant from funding agencies in the public, commercial, or not-for-profit sectors.
We thank Stefano Iubini for a careful reading of the manuscript. 

\section*{Appendix}
We give here some basic details of the multiple-scale expansion leading
to the effective equations. Derivatives are expanded at different orders
\begin{eqnarray}
&O(1):& \dd{}\dd{V_n}\\
&O(\varepsilon):&  2\dd{}\DD{V_n}+\dd{}\dd{\V_n}\\
&O(\varepsilon^2):& \DD{}\DD{V_n}+2\dd{}\DD{\V_n}+\dd{}\dd{\VV_n}
\end{eqnarray}
Similar expansions hold for the variables $U$. It is matter 
to solve the resulting equation order-by-order.
We then assume an expansion of the force as $F(0)=0$, $F'(0)=m\omega_0^2$ 
$$
F(x) = -m\omega_0^2 x - \alpha x^2 -\beta x^3 \ldots = -m\omega_0^2 x +
\varepsilon \F(x)
$$
As far as smallness of nonlinearity is concerned, we have here 
assumed that $\alpha$ and $\beta$ are of the same order of $\varepsilon$
but one may as well think of $\beta\sim \varepsilon^2$ 
in this case $\beta$ does not enter up to second order.
To zeroth order $\varepsilon=0$ the problem 
reduces to uncoupled linear oscillators 
\begin{eqnarray}
&& \dd{}\dd{V_n}= \omega_0^2(U_n+U_{n-1}-2V_n)\nonumber\\
&& \dd{}\dd{U_n}=  0
\end{eqnarray}
We can thus take $U_n(t,T)$ to be independent of $t$ and solve the above as  
\begin{equation}
V_n(t,T)=  \frac {U_n+U_{n-1}}{2} + A_n(T)  e^{i\Omega_0 t} + c.c.
\label{ord0}
\end{equation}
where $\Omega_0 = \sqrt{2}\omega_0$ the upper band edge of the optical
band. The interpretation is simply that  oscillations of the 
light particles are decoupled and occur 
around the center of mass of the neighboring heavy particles
with a slow modulation given by the complex amplitudes $A_n(T)$.


To first order in $\varepsilon$
\begin{eqnarray*}
&& 2\dd{}\DD{V_n}+\dd{}\dd{\V_n}= \omega_0^2(\U_n+\U_{n-1}-2\V_n)
- \F(U_{n} - V_{n}) + \F( V_{n} - U_{n-1})\\
&& 2\dd{}\DD{U_n}+\dd{}\dd{\U_n}= 0
\end{eqnarray*}
Taking into account the zeroth order, we get 
\begin{eqnarray}
&&U_n-V_n=   \frac {U_n-U_{n-1}}{2} - A_n(T)  e^{i\Omega_0 t} + c.c.
\nonumber\\
&&V_n-U_{n-1} = \frac {U_n-U_{n-1}}{2} + A_n(T)  e^{i\Omega_0 t} + c.c.
\label{V}
\end{eqnarray}
That can be substituted in the equations above to yield
\begin{eqnarray*}
&& \dd{}\dd{\V_n}-\omega_0^2(\U_n+\U_{n-1}-2\V_n)=
-2\dd{}\DD{V_n}  - \F(U_{n} - V_{n}) + \F( V_{n} - U_{n-1}) \\
&& \dd{}\dd{\U_n} = 0.
\end{eqnarray*}
The second equation implies $\U_n(T)$ so that $\V_n$ oscillates with frequency
$\Omega_0$ on the fast scale.
To avoid secular terms, one has to impose that the right hand
side in the first equation does not contain terms in 
$e^{\pm i\Omega_0 t}$, obtaining the condition 
\begin{equation}
2 i\Omega_0 \DD{A_n}= -3\beta |A_n|^2A_n 
-2\alpha\left(U_n-U_{n-1}\right)A_n
\end{equation}
which has the familiar form of the amplitude equation for a nonlinear
oscillator except for the explicit dependence on the variables $U_n$ 
that, to this order are yet to be determined.
Actually, the equation for $\V$ is nonhomogeneous
\begin{equation}
\dd{}\dd{\V_n}-\omega_0^2(\U_n+\U_{n-1}-2\V_n)=
-2\beta A_n^3 e^{3i\Omega_0t} + cc
\label{3omega}
\end{equation}
which would requiring considering the third harmonic of 
the nonlinear oscillations. To keep things simpler we 
neglect at all the rhs of eq. (\ref{3omega}) so that
\begin{equation}
\V_n(t,T)=  \frac {\U_n+\U_{n-1}}{2} + \A_n(T)  e^{i\Omega_0 t} + c.c. 
\label{ord1}
\end{equation}


To second order in $\varepsilon$
\begin{eqnarray*}
&& \DD{}\DD{V_n}+2\dd{}\DD{\V_n}+\dd{}\dd{\VV_n}= \\
&& \omega_0^2(\UU_n+\UU_{n-1}-2\VV_n)
- \F'(U_{n} - V_{n})(\U_{n} - \V_{n})
 + \F'( V_{n} - U_{n-1})( \V_{n} - \U_{n-1})\\
&& \DD{}\DD{U_n}+2\dd{}\DD{\U_n}+\dd{}\dd{\UU_n}= 
\omega_0^2[V_{n+1} - 2U_{n} + V_{n}]
\end{eqnarray*}
Using previous orders, eq.(\ref{V})
\begin{eqnarray}
&& \dd{}\dd{\VV_n}-\omega_0^2(\UU_n+\UU_{n-1}-2\VV_n)= \nonumber\\
&&  =-\DD{}\DD{V_n}-2\dd{}\DD{\V_n}
- \F'(U_{n} - V_{n})(\U_{n} - \V_{n})
 + \F'( V_{n} - U_{n-1})( \V_{n} - \U_{n-1})
 \nonumber\\
&& \dd{}\dd{\UU_n}= -\DD{}\DD{U_n}+
\frac12\omega_0^2[U_{n+1} - 2U_{n} + U_{n-1}]+
(A_{n+1}+A_n)e^{i\Omega_0t} + cc
\label{ord2}
\end{eqnarray}
A solution of the second equation above is 
\begin{eqnarray}
\DD{}\DD{U_n}=
\frac12\omega_0^2[U_{n+1} - 2U_{n} + U_{n-1}] \\
\UU_n(t,T) = -\frac{A_{n+1}+A_n}{\Omega_0^2} \,e^{i\Omega_0 t} + c.c.
\label{u2}
\end{eqnarray}
This shows that the heavy masses have also a fast component
to order $\varepsilon^2$.
Once more, to avoid secular term we must impose 
that terms in $e^{\pm i\Omega_0 t}$ in the first of eqs.(\ref{ord2})
are identically zero yielding
\begin{equation}
2 i\Omega_0 \DD{\A_n}= -\DD{}\DD{A_n} -2\alpha\left[(U_n-U_{n-1})\A_n+
(\U_n-\U_{n-1})A_n
\right]+\omega_0^2(\UU_n+\UU_{n-1})
\end{equation}
Note that the terms in $\beta$ do not enter.
To close the system we need an equation for $\U_n$ which however seems
undetermined at least at this order (probably this would require a 
$\varepsilon^3$ terms). If we suppose  $\U_n=0$, using eq. (\ref{u2})
we have the equation for $\A_n$
\begin{eqnarray}
2 i\Omega_0 \DD{\A_n}= -\DD{}\DD{A_n} -2\alpha(U_n-U_{n-1})\A_n
-\frac{1}{2}(A_{n+1}+2A_n+A_{n-1}).
\label{a1}
\end{eqnarray} 
Finally, we can write
a closed set of equations for $U_n+\varepsilon^2\UU_n$ 
and the modulation $\mathcal{A}_n=A_n+\varepsilon \A_n $ summing the corresponding equations for first and second order.
To obtain the dissipation term as in (\ref{oscmod}),  in (\ref{a1})  we 
make the replacement
\[
\DD{}\DD{A_n} \approx i\Omega_0 \DD{A_n}
\]
which is justified within the slowly-variable amplitude hypotesis.
The simplified equations (\ref{oscmod}) are obtained
neglecting $\UU_n$ and the last term in (\ref{a1}).


\bibliography{dbdiatomic}{}

\end{document}